\documentclass[twocolumn,prl,letterpaper,showpacs]{revtex4}
\usepackage{amsmath}
\usepackage{graphicx}
\usepackage{amssymb}
\usepackage{color}
\usepackage{float}

\begin{document}


\title{Emission of non-classical radiation by inelastic Cooper pair tunneling}

\author{M. \surname{Westig}$^{1}$}

\author{B. \surname{Kubala}$^{2}$}

\author{O. \surname{Parlavecchio}$^{1}$}

\author{Y. \surname{Mukharsky}$^{1}$}

\author{C. \surname{Altimiras}$^{1}$}

\author{P. \surname{Joyez}$^{1}$}

\author{D. \surname{Vion}$^{1}$}

\author{P. \surname{Roche}$^{1}$}

\author{D. \surname{Esteve}$^{1}$}

\author{M. \surname{Hofheinz}$^{1}$}
\altaffiliation[Present address: ]{CEA, INAC-SPSMS, F-38000 Grenoble, France}

\author{M. \surname{Trif}$^{3}$}

\author{P. \surname{Simon}$^{3}$}

\author{J. \surname{Ankerhold}$^{2}$}
\email{email: joachim.ankerhold@uni-ulm.de}

\author{F. \surname{Portier}$^{1}$}
\email{email: fabien.portier@cea.fr}

\affiliation{$^{1}$ SPEC (UMR 3680 CEA-CNRS), CEA Saclay, 91191 Gif-sur-Yvette,
France}
\affiliation{$^{2}$ Institute for Complex Quantum Systems and Institute for Complex Quantum Systems, University of Ulm, 89069 Ulm, Germany}
\affiliation{$^{3}$ Laboratoire de Physique des Solides, Universit\'e Paris-Sud, 91405 Orsay, France}

\begin{abstract}
We show that a properly dc-biased Josephson junction in series with two microwave resonators of different frequencies emits photon pairs in the resonators. By measuring auto- and inter-correlations of the power leaking out of the resonators, we demonstrate two-mode amplitude squeezing below the classical limit. This non-classical microwave light emission is found to be in quantitative agreement with our theoretical predictions, up to an emission rate of 2 billion photon pairs per second. 
\end{abstract}

\date{\today}

\maketitle

Microwave radiation is usually produced by ac-driving a conductor like a
wire antenna. The radiated field is then a so-called coherent state \cite{PhysRev.131.2766} which
closely resembles a classical state. On the other hand, a simply dc-biased
quantum conductor can also generate microwave radiation, owing to the
probabilistic nature of the discrete charge transfer through the conductor
which cause quantum fluctuations of the current \cite{PhysRevLett.78.3370, PhysRevLett.99.236803, PhysRevLett.111.136601}. In this latter situation, it is expected that the quantum character of the charge transfer
may imprint in the properties of the emitted radiation, possibly leading to nonclassical
radiation, such as e.g. antibunched photons \cite{PhysRevLett.86.700,PhysRevLett.93.096801,PhysRevB.81.115331,PhysRevB.81.155421,PhysRevB.92.195417, PhysRevLett.111.247002, PhysRevLett.115.027004}. More broadly, one
may wonder what other types of interesting or useful nonclassical states of light can be generated with such a simple method. In this Letter, we investigate the properties of photons pairs emitted by a dc voltage-biased Josephson junction. In such a junction, at bias voltage less than the gap voltage $2 \Delta/ e$, no quasiparticle excitation can be created in the superconducting electrodes. Thus, a DC current can only flow through the junction when the electrostatic energy $2 e V$ associated to transfer of the charge of a Cooper pair through the circuit is absorbed by modes of the surrounding
circuit \cite{averin90,ingold92,holst94,basset10,PhysRevLett.106.217005,PhysRevB.85.085435}.

\begin{figure}
\centering
\includegraphics[width= 8.2cm]{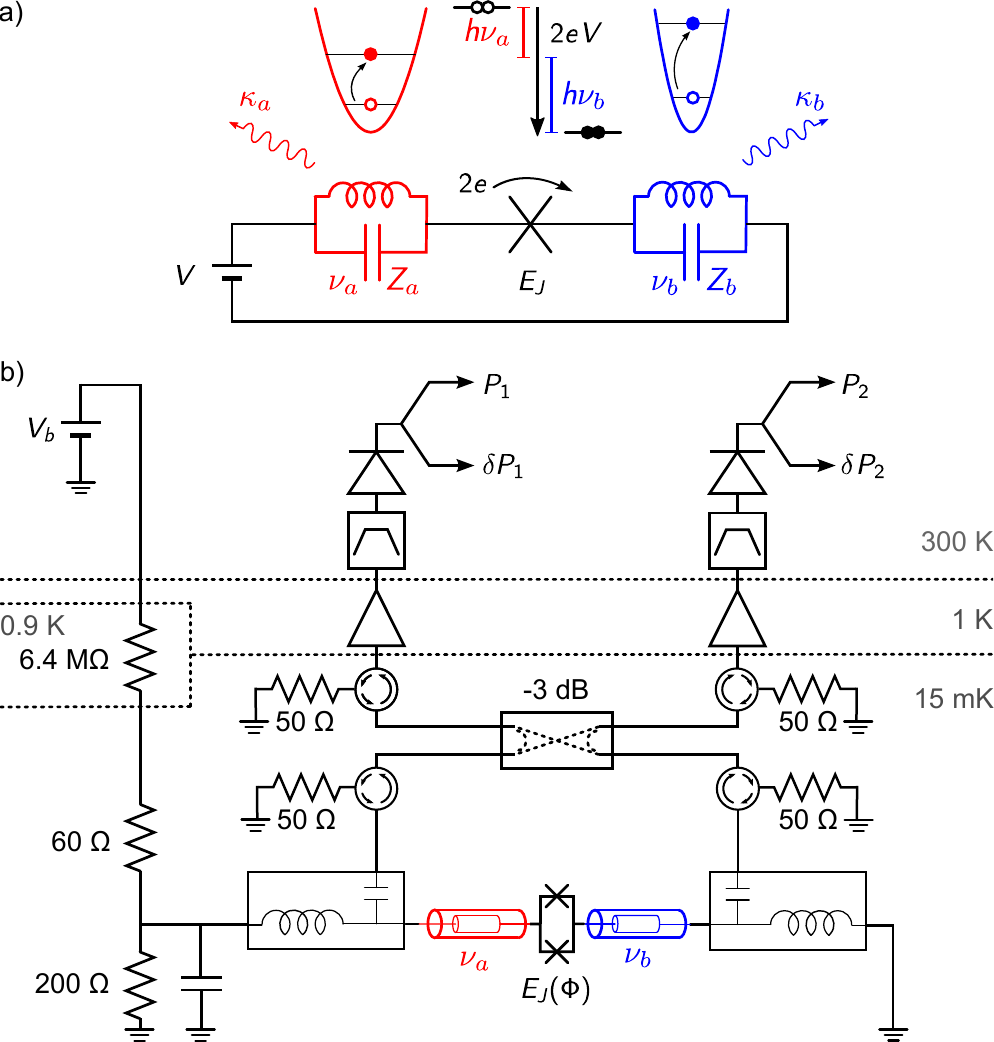}
\caption{\label{fig:scheme}
\textbf{Principle and setup of the experiment.} (a)
a Josephson junction in series with two resonators with frequencies $\nu_{a,b}$ emits a photon pair in the resonators each time a Cooper pair tunnels through it at a dc bias voltage $V$ such that $2eV=h\nu_a+h\nu_b$. Microwave radiations leaking out of the resonators at rates $\kappa_{a,b}$ should present strong quantum correlations.  (b) Setup: The
sample consists of a SQUID working as a magnetically tunable Josephson
junction, in series with two  3-quarterwave
transformer resonators. Two bias tees make possible to dc voltage bias the SQUID while collecting radiation  from the resonators. A Hanbury-Brown and Twiss setup with a hybrid coupler, isolators, amplifiers, filters, and power detectors is used to measure all powers and power-power correlations (see text).}
\end{figure}

In order to obtain a situation in which the quantum nature of the emitted radiation can be probed quantitatively, we place such a dc-biased Josephson junction in an engineered environment made of two series resonators with different frequencies $\nu_a$, $\nu_b$, as shown in Fig.~\ref{fig:scheme}a. We consider in particular the resonance condition $2eV= h (\nu_a+ \nu_b)$, at which the transfer of a single Cooper pair is expected to create one photon in each resonator, leaking afterwards in two microwave lines. By measuring both photon emission rates as well as the power-power auto and intercorrelations, we prove that these correlations violate a Cauchy-Schwartz inequality obeyed by classical light, meaning that the relative fluctuations of the outgoing modes are suppressed below the classical limit. This two-mode amplitude squeezing is observed for emission rates as high as $2 \times 10^9$ photon pairs per second, making our setup a particularly bright (and simple) source of nonclassical radiation.

Our experimental setup is shown in Fig.~\ref{fig:scheme}b: a small superconducting quantum interference device 
(SQUID) acts as a tunable Josephson
junction with Josephson energy $E_{{\mathrm J}}=E_{{\mathrm J} 0}|\cos(2e\Phi/\hbar)|$ adjustable via the
magnetic flux $\Phi$ threading its loop. The two resonators connected to either side of the SQUID
are made of three cascaded quarter-wave transformers.
Their expected fundamental modes have frequencies  $\nu_{a, b}\simeq 5.1, 7.0~\mathrm{GHz}$ and characteristic impedances $Z_{a, b}
\simeq 140~\mathrm{\Omega}$. The resonators are connected to two separate bias tees making it possible to dc voltage bias the SQUID while collecting radiation on two separate microwave lines with wave impedance 50$\mathrm{\Omega}$. The resonator quality factors $Q_{a,b} \simeq 25, 35$ are thus determined by the energy leaking rate $\kappa_{a,b} \simeq 1.3 \  10^9~\mathrm{s}^{-1}$ into each microwave line. The expected total series impedance $Z(\nu)$ seen by the SQUID thus reaches $\simeq 3.2, 4.9~\mathrm{k\Omega}$ for modes $a$ and $b$. The two measurement lines are arranged in a Hanbury-Brown and Twiss (HBT) microwave-setup to probe the quantum fluctuations of the emitted radiation without being blinded by the noise of the amplification chains: they are connected through two isolators to a $90^\circ$ hybrid coupler acting as a microwave beam splitter. The two lines after the coupler (hereafter called 1 and 2) thus propagate half of the powers leaking from resonators $a$ and $b$. The two outputs of the beam splitter are sent through two additional isolators and filters to two microwave high-electron-mobility transistor (HEMT) amplifiers  placed at 4.2 K. These isolators and filters protect the sample from the amplifiers' back-action noise and ensure thermalization of its environment during the experiment. They also attenuate the signals by about 3 dB. After further amplification at room temperature (not shown in Fig. \ref{fig:scheme}-b), the signals are filtered either by a heterodyne technique implementing a 12 MHz-wide band pass filter at tunable frequency 
or by adjustable bandpass cavity filters covering only one of the resonator lines. In both cases, the filtered signal is detected by a quadratic detector, whose output voltage 
is proportional to its input ac power $P_{1,2}(t)$. In order to extract the small average contribution $\left\langle P_{1,2}^S\right\rangle$ of the sample from the 
large background noise of the cryogenic amplifiers, we apply a $0$ to $V$ square-wave modulation at 113 Hz to the sample 
bias and perform a lock-in detection of the square-wave response of the quadratic detectors.

\begin{figure*}
\centering
\includegraphics[width=17cm]{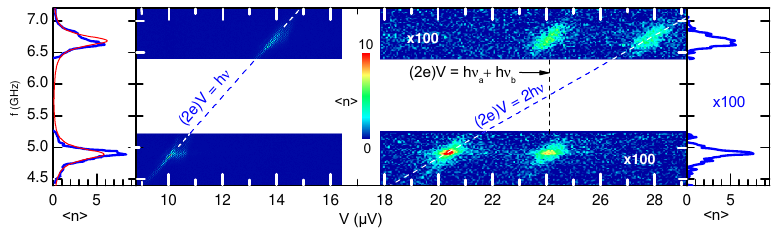}
\caption{\label{fig:peakDenis}
Detected power spectral density as a function of frequency $\nu$ and dc bias voltage $V$, re-expressed in terms of photon rate per unit bandwidth $\langle n \rangle$ at the output of the resonators. Emission
occurs at one photon per Cooper pair along the line $2 eV = h \nu$ (see left 2D map), and  at two identical photons per Cooper pair along the $2eV=2 h\nu$ line and one pair of photons in modes $a,b$ along the vertical line $2eV=h\nu_a+h\nu_b$. (see right 2D map). The leftmost panel is a cut along the $2ev=h\nu$ line showing that the emitted power (blue bold line) follows the $\mathrm{Re}Z(\nu)/\nu$ prediction (red thin line) of the $P(E)$-theory with $E_{{\mathrm J}}$ being the only adjustable parameter. The rightmost panel is a cut at $V=24.1 \mu \mathrm{V}=h(\nu_a+\nu_b)/(2e)$.
Occupation numbers for the two photons processes have been multiplied by 100 for clarity.}
\end{figure*}

The sample is cooled to 15 mK in a dilution refrigerator. We first characterize in-situ our sample and detection chain using the quasi-particle shot noise as a calibrated source \cite{PhysRevLett.106.217005}: We measure the power emitted by the SQUID at bias voltage $V\simeq 0.975~\mathrm{mV}$, well above the gap voltage $2\Delta/e \simeq 0.4~\mathrm{mV}$.
Under these conditions, the voltage derivative of the measured power spectral density reads
$2e \mathrm{Re} Z(\nu) R_n/ |R_n+Z(\nu)|^{2} \times G$ with $R_n = 8.0~\mathrm{k\Omega}$
 the tunnel resistance of the SQUID in the normal state, and $G$ the total gain of the setup.
The measured frequency dependence is in good agreement with the above formula, using our design of $Z(\nu)$. This measurement thus provides an in-situ determination of gain $G$.  More information on the design and comparison with the high bias data can be found in the supplementary material \cite{SM}.

We then measure the photon emission rate as a function of frequency and  bias voltage for the single photon (left side of fig.~\ref{fig:peakDenis}) and two photon emission processes (right side of fig.~\ref{fig:peakDenis}). To do so, we ensure a maximum population of the resonators of order unity by setting $E_{{\mathrm J}}$ at a sufficiently small value \cite{NoteEJ}. The single photon processes occur along the $2eV=h\nu$ line, with an intensity modulated by $\mathrm{Re} Z(\nu)$. At fixed bias voltage, the 12~MHz spectral width of the detected radiation coincides with our detection bandwidth, proving that our bias line is well filtered and adds to the Josephson frequency $2 eV/h$ an incertitude negligible compared with the width of the resonators $\kappa_{a,b}/2\pi$. Fainter lines appear at $2eV=h(\nu_{a,b} \pm m \nu_P)$, with $\nu_P= 35~\mathrm{MHz}$ and $m$ an integer. We attribute these satellite peaks to the existence of a parasitic resonance at frequency $\nu_P$, allowing for multi-photon processes with one photon emitted at high frequency $\nu_{a}$ or $\nu_{b}$ and $m$ photons emitted into or absorbed from the parasitic mode. From the relative weight of the peaks (data not shown here), we estimate the impedance of this parasitic mode to $Z_P= 204~\Omega$, with a thermal population of $n_p\sim 8$ photons corresponding to a 14.5 mK temperature for the $\nu_P$ mode. This is in good agreement with the measured
fridge temperature of 15 mK $\pm$ 1mK.

At higher bias voltages (right side of Fig.~\ref{fig:peakDenis}) we detect processes for which the tunneling of a Cooper pair is associated to the emission of two photons: At $V= 20.2~\mu$V (resp.~27.9$~\mu$V), we detect radiation around the frequency of resonator $a$ (resp.~$b$) due to the simultaneous emission of two photons into this resonator for each Cooper pair tunneling through the junction. At an intermediate voltage $V=24.1~\mu$V, we detect radiation at both frequencies $\nu_a$ and $\nu_b$, due to the simultaneous emission of a photon in each resonator for each Cooper pair transferred. The rightmost panel of Fig.~\ref{fig:peakDenis} shows the corresponding power spectral density of the emitted radiation. Integrating this spectral power over a 500 MHz bandwidth centered around $\nu_a$ and $\nu_b$ indeed shows that the photon emission rates into resonators $a$ and $b$ coincide (within 10\% due to calibration uncertainties). At fixed bias voltage $V$ the spectral width of the emitted radiation from any two photon process is comparable with the width of the resonators: Due to energy conservation, the sum of the frequencies of the two emitted photons is equal to the Josephson frequency $\nu_J=2eV/h$. As a consequence, if one of the photons is emitted at frequency $\nu$, the other is emitted at frequency $\nu_J-\nu$. The corresponding weight is given by the product of the environment's impedances $\mathrm{Re}[Z(\nu)] \mathrm{Re}[Z(\nu_J-\nu)]/R_Q^2$, resulting in a width of the emitted radiation of the order of half of the resonator's width \cite{PhysRevLett.106.217005}.

It is quite intuitive that a common excitation process that creates one photon in each resonator for each Cooper pair tunneling  through the junction yields strong non-classical correlations of the resonators' occupation numbers $n_a=a^\dagger a$ and $n_b=b^\dagger b$. This effect is quantified by the so-called noise reduction factor $\mathrm{NRF}=\mathrm{Var}(n_a-n_b)/\left\langle n_a+n_b\right\rangle$, i.e.~the variance of the occupation difference, normalized to the average total number of photons, yielding 1 in the case of two independent coherent states. With photon pair creation in non-leaking resonators, $n_a$ and $n_b$ would remain equal and NRF would be reduced to zero. In reality, due to the uncorrelated energy decays of the two resonators, $n_a$ and $n_b$ do not remain equal, even for perfectly symmetric modes, and NRF is expected to increase from 0 to 1/2  \cite{PhysRevB.91.184508,PhysRevB.92.014503}.

The NRF can be linked to the zero-delay value of second order coherence functions
\begin{equation*}
g^{(2)}_{\alpha,\beta}(\tau)= \frac{\left\langle \alpha^\dagger(\tau)\beta^\dagger(0)\alpha(\tau)\beta(0)\right\rangle}{\left\langle \alpha^\dagger(\tau)\alpha(\tau)\right\rangle\left\langle \beta^\dagger(0)\beta(0)\right\rangle}\, 
\end{equation*}
with $\alpha,\beta\in \left\{a,b\right\}$. We get 

\begin{equation}
\begin{split}
\mathrm{NRF}&=1+\frac{\left\langle n_a\right\rangle^2 g^{(2)}_{a,a}(0)+\left\langle n_b\right\rangle^2 g^{(2)}_{b,b}(0)-2 \left\langle n_a\right\rangle \left\langle n_b\right\rangle g^{(2)}_{a,b}(0)}{\left\langle n_a+n_b\right\rangle} 
\\ &=1+\left\langle n\right\rangle \frac{g^{(2)}_{a,a}(0)+ g^{(2)}_{b,b}(0)-2 g^{(2)}_{a,b}(0)}{2} 
\end{split}
\label{Eq.NRF}
\end{equation}

\noindent for $\left\langle n_a\right\rangle=\left\langle n_b\right\rangle=\left\langle n\right\rangle$. A classical bound $\mathrm{NRF}\geq 1$ follows from the Cauchy-Schwarz inequality
\begin{equation}
g^{(2)}_{a,b}(0)\le \frac{g^{(2)}_{a,a}(0)+g^{(2)}_{b,b}(0)}{2},\
\label{CS}
\end{equation}

\noindent valid for two classical fields, i.e. for a two-mode density operator corresponding to any statistical mixture of coherent states.  It is easy to explain why the above inequality must be violated in our situation, with hence a
NRF below 1: for low Cooper pair tunneling rates, photons have time to leak out of the resonators between each $a^\dagger b^\dagger$ photon pair creation events. The probabilities to simultaneously find two photons in the same mode, as measured by the auto-correlation $g^{(2)}_{\alpha \alpha}(0)$ is then close to zero while the cross-correlation $g^{(2)}_{ab}(0)$ giving the probability to find simultaneously one photon in each mode is high \cite{PhysRevLett.110.267004}. This situation corresponds to a squeezing of the relative amplitudes of the two modes below the classical limit \cite{0954-8998-1-1-001, PhysRevLett.59.2555, PhysRevLett.102.213602, PhysRevLett.113.043602}.

\begin{figure}[t]
\centering
\includegraphics[width= 9 cm]{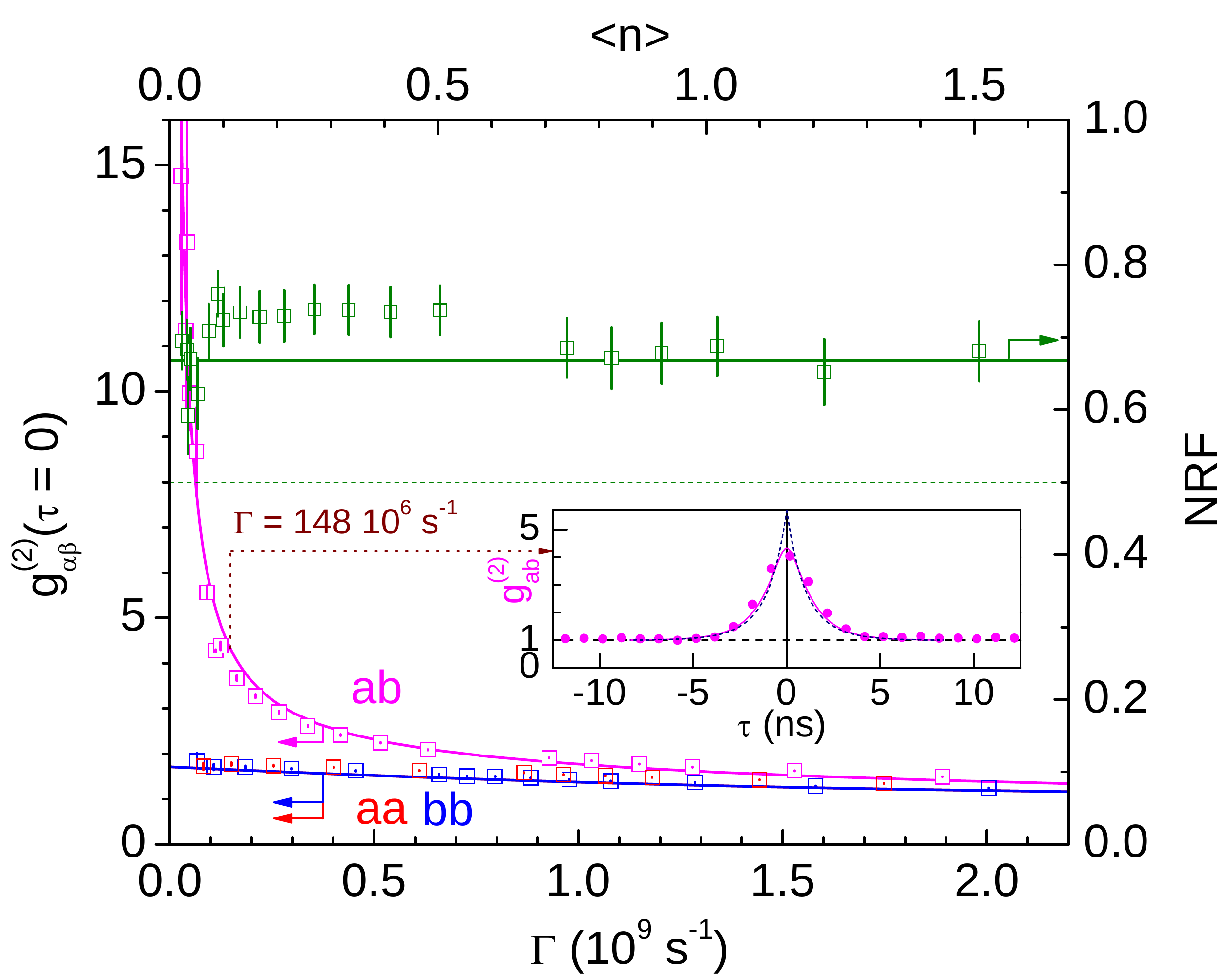}
\caption{\label{fig:g2NRF}
\textbf{Non classicality of the emitted radiation} at bias $V = 24.1 \mu$V, as a function of the photon pair emission rate.
Left scale: Zero delay power-power correlation functions $g^{(2)}_{aa}(0)$ (red open squares), $g^{(2)}_{ab}(0)$ (magenta open squares) and $g^{(2)}_{bb}(0)$
(blue open squares). Right scale: the corresponding
NRF (green open squares) extracted from  Eq. \ref{Eq.NRF} does not reach the ideal value of 0.5 (horizontal dashed line), but remains well below 1, which demonstrates two-mode amplitude squeezing.
Inset: Time dependence of $g^{(2)}_{ab}(\tau)$. The solid lines are theoretical predictions without adjustable parameters. In the inset, the magenta curve includes the effect of the detector finite response time, while the dark blue dashed corresponds to the prediction for infinitely fast detectors.}
\end{figure}

To experimentally probe this violation, we collect the photons leaking out into the measurement lines. At the resonator outputs, the three functions $g^{(2)}_{\alpha_L,\beta_L}$, where $\alpha_L =\sqrt{\kappa_{\alpha}} \alpha$ and $\beta_L =\sqrt{\kappa_{\beta}} \beta$ are the propagating field operators, are simply equal to $g^{(2)}_{\alpha,\beta}$ inside the resonators. Both propagating fields $a_L$ and $b_L$ are then beam-splitted and sent to lines 1 and 2, which include $\simeq 650$~MHz-wide filters centered around $\nu_\alpha$ and $\nu_\beta$ to select the desired resonator contributions. Measuring the output powers $P_1(t)$ and $P_2(t)$ using two Herotek DTM 180AA fast quadratic detectors with a $0.42 \pm 0.02$~ns response time \cite{NoteQDtimeres}, we obtain the correlation functions 
\begin{equation}
g^{(2)}_{\alpha,\beta}(\tau)= 1 + \frac{\left\langle \delta P_1(t+\tau)\delta P_2(t)\right\rangle}{\left\langle P_1^S\right\rangle\left\langle P_2^S\right\rangle},
\label{eq:g2exp}
\end{equation}
where $\left\langle P_{1,2}^S\right\rangle=\left\langle P_{1,2}(V,t)\right\rangle-\left\langle P_{1,2}(0,t)\right\rangle$  are the average sample contributions and $\delta P_{1,2}(t)=P_{1,2}(V,t)-\left\langle P_{1,2}(V,t)\right\rangle$ are the power fluctuations. The advantage of this strategy is that it gives access to the fluctuations of the power emitted by the sample avoiding parasitic terms due to the much higher noise power of the HEMT amplifiers \cite{PhysRevLett.93.056801,PhysRevLett.104.206802,PhysRevLett.108.263601}.
Figure \ref{fig:g2NRF} shows the three coherence functions  $g^{(2)}_{\alpha,\beta}$ at zero delay $\tau$ as well as the noise reduction factor $\mathrm{NRF}$,  as a function of the photon pair emission rate $\Gamma$, the later being varied by scanning the flux threading the SQUID loop. The figure shows that inequality (\ref{CS}) is indeed violated for photon emission rates up to $2\times 10^9$ photon pairs per second, the $\mathrm{NRF}$ remaining close to 0.7. The decay of  $g^{(2)}_{a,b}(\tau)$ due to the independent resonator leakage is shown in the inset.

To compare our measurements with theory, we compute the $g^{(2)}_{\alpha,\beta}(\tau)$  functions. This task goes beyond the framework of the standard Dynamical Coulomb Blockade theory, which assumes that the electromagnetic environment of the junction remains in equilibrium. Here instead we need to predict how the presence of photons already emitted in the resonators modifies the next emission process. To do so, one can develop an input-output approach \cite{PhysRevLett.110.267004, JuhaNJP}. Equivalently, we use here a Lindblad master equation approach, starting from the Hamiltonian
\begin{equation}
\begin{split}
H=& h \nu  _a a^{\dagger} a + h \nu_b b^{\dagger} b \\
& - E_J \cos \left[2eV t/\hbar +  \Delta_{a} \left( a^{\dagger} + a \right) +  \Delta_{b} \left( b^{\dagger} + b \right) \right]
\end{split}
\end{equation}

\noindent  modeling the two resonators coupled to the voltage biased-junction $V$ \cite{PhysRevLett.111.247002,PhysRevB.91.184508,PhysRevB.92.014503}, with $\Delta_{a,b}=(\pi Z_{a,b}/R_Q)^{1/2}$ and $R_Q=h/4e^2$. Assuming $2 eV = h (\nu_a+\nu_b)$ and moving to the frame rotating at $\omega_J=2eV/\hbar$, the Hamiltonian in the rotating wave approximation then reads 
\begin{equation} \label{Hamiltonian}
H_{RW} =\frac{{E}_J^*}{2}:\frac{J_1(2\Delta_a\sqrt{a^{\dagger}a})
J_1(2\Delta_b\sqrt{b^{\dagger}b})}{\sqrt{a^{\dagger}a}\sqrt{b^{\dagger}b}}\left(a^{\dagger}b^{\dagger}+ab\right): \end{equation}
\noindent with $E_J^*=E_J{\rm e}^{-(\Delta_a^2+\Delta_b^2)/2}$ the Josephson energy renormalized by the zero-point fluctuations of the two modes, and 
 the colons characters meaning normal ordering of the operators.
 The Bessel functions of the first kind $J_1$  ''dress'' the elementary photon pair creation process $a^{\dagger}b^{\dagger}$ by higher-order corrections in $n_{a,b}$.
Note that for low photon numbers $n_{a,b}$ and for low impedances $Z_{a,b} \ll R_Q$, $H_{RW}$ reduces to 
$H'_{RW}\simeq\frac{{E}_J^*\Delta_a\Delta_b}{2}\left(a^{\dagger}b^{\dagger}+ab\right)\,, \label{eq:ndpa}$
which suffices to qualitatively explain the experimental data.
Photon leakage from the resonators can be accounted for by including damping rates  $\kappa_\alpha$ of standard quantum-optical form (in the $T=0$ limit) in the quantum master equation of the system,
\begin{equation}\label{Lindblad}
\dot{\rho} = -\frac{i}{\hbar}[{H_{RW}},\rho]+\sum_{\alpha=a,b}\kappa_\alpha\left(2 \alpha \rho \alpha^{\dagger}-\alpha^{\dagger}\alpha\rho-\rho \alpha^{\dagger}\alpha\right).
\end{equation}
 Additional incoherent dynamics of the $a$ and $b$ modes is caused by the parasitic low frequency mode $\nu_P$ \cite{PhysRevLett.111.247002} and broadens the one-photon resonances. However, we find that it has little impact on the two photon $a$-$b$ resonance.

Simulating  (\ref{Lindblad}) yields $\rho(t)$ and hence all $g^{(2)}_{\alpha \beta}(\tau)$ functions. These functions, convoluted with the $0.42 \pm 0.02$~ns detector response mentioned above, are plotted as lines in Fig.~\ref{fig:g2NRF}. They are found in agreement with the experimental results. Note that the deviation from NRF=1/2 seen in Fig.~\ref{fig:g2NRF} is almost only due to this finite response time.  

In conclusion we have shown that a DC-biased Josephson junction in series with two resonators provides a simple and bright source of non-classical radiation, displaying relative fluctuations of the populations of the two modes below the classical limit. We have also presented a theory which  quantitatively  accounts for our experimental findings. While the present experiment is performed at microwave frequencies using Aluminum Josephson junctions, the physics involved here can be transposed to higher gap superconductors, such as NbTiN or even YBaCuO, opening the possibility of creating non classical THz radiations.  We gratefully acknowledge partial support from “Investissements d’Avenir” LabEx PALM (ANR-10-LABX-0039-PALM), ANR contracts ANPhoTeQ and GEARED, and from the ERC through the NSECPROBE grant.

		

\clearpage

\onecolumngrid
\vspace{\columnsep}
\begin{center}
\textbf{\large Emission of non classical radiation by inelastic Cooper pair tunneling:\\Supplemental material}
\end{center}
\vspace{\columnsep}
\twocolumngrid

\setcounter{equation}{0}
\setcounter{figure}{0}
\setcounter{table}{0}
\setcounter{page}{1}
\makeatletter
\renewcommand{\theequation}{S\arabic{equation}}
\renewcommand{\thefigure}{S\arabic{figure}}

\newcommand\id{\ensuremath{\mathbbm{1}}}



\section*{General predictions for $g^{(2)}$} 
Choosing a particular dc-bias voltage applied to the Josephson junction, we picked out that resonance, where each tunneling Cooper pair excites two photons, one in each resonator. In the steady state this common excitation process is balanced by the decay of photons from the cavity, so that the rate of excitation by tunneling Cooper pairs, $I_\mathrm{dc}/2e$, and the photon leakage rates match
\begin{equation}
\frac{I_\mathrm{dc}}{2e}=\kappa_a \langle n_a \rangle =\kappa_b \langle n_b \rangle\;.
\end{equation}
For equal cavity damping the mean occupations, $\langle n_\alpha \rangle $, are thus identical.

This can be exploited to derive a relation \cite{PhysRevB.91.184508} between cross- and autocorrelations for arbitrary driving strength,
\begin{equation}\label{eq:gab}
g^{(2)}_{ab}(0)=\frac{1}{2 n} +\frac{1}{2}\left[g^{(2)}_{aa}(0)+g^{(2)}_{bb}(0)\right]\, .
\end{equation}
It implies a violation of the classical Cauchy-Schwarz inequality and predicts a noise reduction factor ${\rm NRF}\equiv \frac{1}{2}$, independent of the driving strength and the impedance parameters. 
In the weak driving regime one furthermore finds $g^{(2)}_{ab}(\tau=0) \approx 1/(2\langle n_\alpha\rangle)$, while $g^{(2)}_{\alpha \alpha}(0)=2$.

For asymmetric cavity damping, $\kappa_a/\kappa_b=r^2 \neq  1$, we instead find
\begin{equation}
g^{(2)}_{ab}(0)=\frac{r}{1+r^2} \left[ \frac{1}{\sqrt{\langle n_a \rangle \langle n_b \rangle }} + \frac{1}{r}g^{(2)}_{aa}(0)+rg^{(2)}_{bb}(0)\right],
\end{equation}
resulting in 
\begin{equation}
\mathrm{NRF} = \frac{1+r^4}{(1+r^2)^2}
\end{equation}
in the weak driving limit.
For our experiment, where the decay rates are found to be identical within about 10{\%}, the maximal deviation of the NRF from the value of 1/2  taken for perfectly symmetric rates is well below a percent.

\section*{Basic rate-equation model} 
To pinpoint the physical ingredients necessary for understanding and explaining the essential results presented in  this paper, we set up a simple rate equation model. It allows to reproduce most of the important results in the weak-driving limit of the full theory and qualitatively describes the experimental data in that regime. 

\begin{figure}[tb]
\centering
\includegraphics[width= 7 cm]{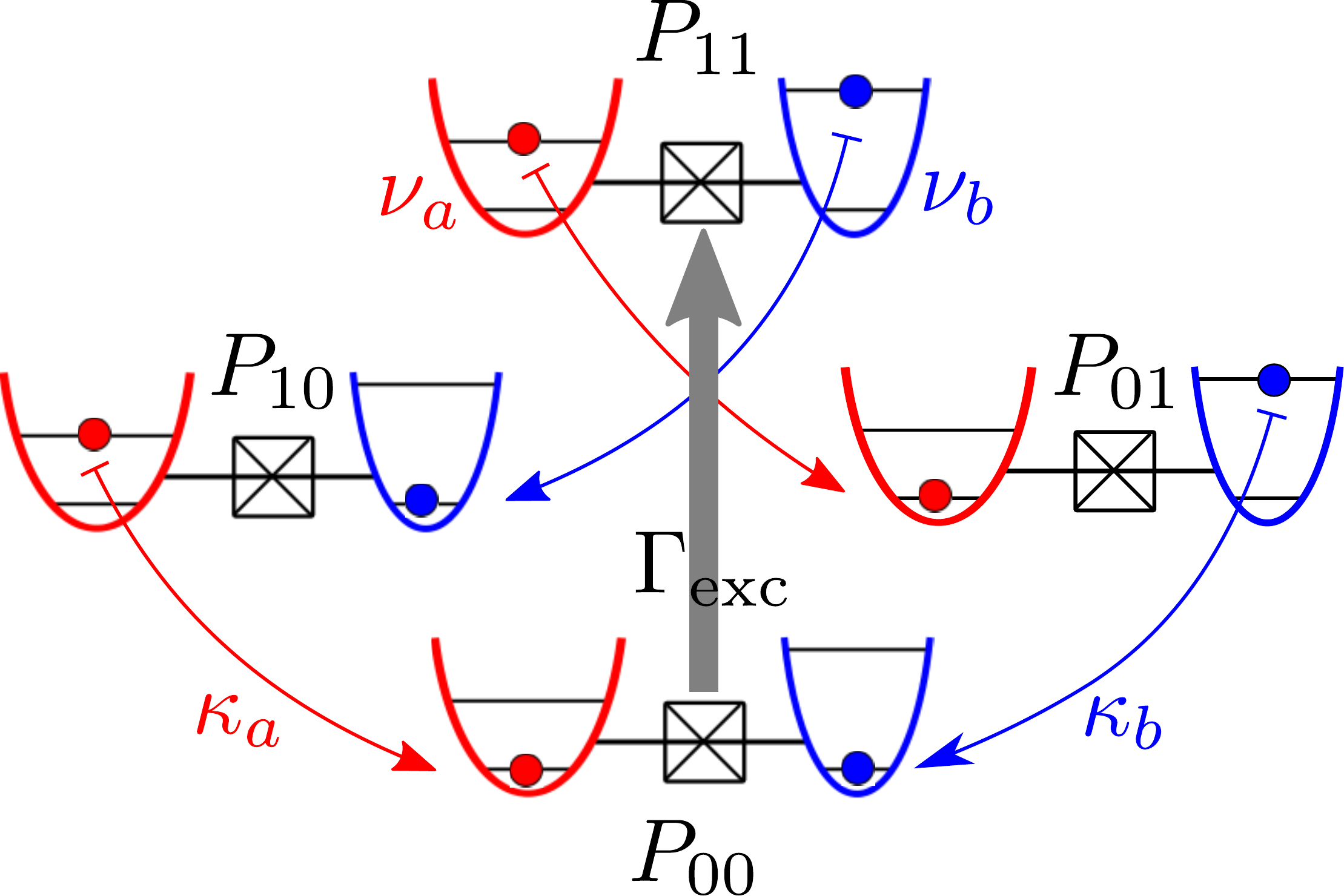}
\caption{\label{fig:rates}
Sketch of a a simple 4-state rate model, which captures essentials of our results for cross-correlations and NRF. 
}
\end{figure}

In the weak driving limit, the results for the cross-correlations and NRF can be found from a simple 4-state rate model, see 
Fig.~\ref{fig:rates}.  A common two-photon process excites the system from the empty state $|n_a=0,\,n_b=0\rangle$ to state $|n_a=1,\,n_b=1\rangle$ with some excitation rate $\Gamma_\mathrm{exc} \propto E_J^2$. Leakage of  photon a (b) with rate $\kappa_{a(b)}$ from the excited state leads to an intermediate state, where cavity a(b) is empty, and finally back to the ground state   $|n_a=0,\,n_b=0\rangle$. Solving for the stationary state of the resulting rate equation, we find that for weak driving, $\Gamma_\mathrm{exc} \ll \kappa_\alpha$ the probability to stay in the ground state remains close to one and all other occupations are of order  $E_J^2$, namely $P_{11} = \Gamma_\mathrm{exc}/(\kappa_a + \kappa_b) = r^2 P_{10} = P_{01}/r^2$. Occupations of higher states are of higher order in  $E_J$, which justifies disregarding them in the 4-state model. 

Cross-correlation and NRF in the weak-driving limit can then be evaluated from these probabilities:
 \begin{equation}
 \begin{aligned}
g^{(2)}_{ab}(0) = \frac{ \langle  n_a n_b\rangle  }{ \langle n_a \rangle \langle n_b \rangle} &= \frac{ P_{11} }{ (P_{11} + P_{10}) (P_{11} + P_{01})   } \\&= \frac{r}{1+r^2} \frac{1}{ \sqrt{ \langle n_a \rangle \langle n_b \rangle }}\;.
\end{aligned}
\end{equation}
The noise reduction factor reduces  to
\begin{equation}
\begin{aligned}
\mathrm{NRF} &=\frac{ \langle (n_a - n_b)^2 \rangle -  \langle n_a - n_b \rangle^2  }{ \langle n_a  n_b \rangle  } \\
&\approx \frac{P_{10}+P_{01}}{P_{10} + P_{01} + 2 P_{11}  } = \frac{1+r^4}{(1+r^2)^2} 
\end{aligned}
\end{equation}
as found above.

The success of the simple rate model highlights the fact, that it is the mere existence of a dominant two-photon excitation process which is at the heart of 
the observed non-classicality. The actual rate drops out of the final results, so that the impedance parameters do not influence the NRF. This also already suggests, that the coherence properties of the process are not relevant and, in fact, we found numerically nearly no impact of low-frequency noise. 

Note, that a proper description of auto-correlations necessarily includes higher occupations and is, hence, beyond the 4-state model. The simplest intuitive result for  $g^{(2)}_{\alpha \alpha}(0)$ is available in the parametric oscillator limit, $\Delta_\alpha \ll1$, where  the enhanced $g^{(2)}_{\alpha \alpha}(0) \approx 2$ can be traced backed to an increased probability for excitation from a state, where there is already some occupation. Technically, this stimulated-emission like enhancement is caused by the corresponding transition-matrix elements entering the tunneling rates.  

~

\section*{Dynamics of the correlation functions} 
The dependence of the second-order coherence functions, $g^{(2)}_{\alpha \beta}(\tau)$, on the delay time $\tau$ is relevant for the results of this paper in order to understand the effects of the detector response on the quantity measured as  $g^{(2)}_{\alpha \beta}(\tau=0)$. 
We find, that convoluting  the theoretically calculated  $g^{(2)}_{\alpha \beta}(\tau)$ with the detector response yields a substantial reduction of the cross-correlations, while the auto-correlations are less affected. As a consequence, the so-predicted reduction of the noise is less pronounced and nicely matches the measurement results of $\mathrm{NRF}\approx 0.7$.

For some intuitive insights, we again consider weak driving, so that the individual two-photon creation processes are well separated in time.
The effective time-averaging caused by the detector affects the NRF only because the origin of auto- and cross-correlations differ, and consequently so do their time-dependences.
 
\emph{Cross-correlations} at  $\tau=0$ stem from a single two-photon creation process. The resulting $1/(2 n)$ contribution decays with the typical lifetime of such an excitation as $\exp{(-\kappa_{\alpha} \tau)}$.

\emph{Auto-correlations} in contrast, do not get a direct contribution from a single tunneling event. In fact, $g^{(2)}_{\alpha\alpha}(\tau =0)$ , i.e. the detection of two-photons at once necessarily involves double occupation of cavity $\alpha$. (As explained above, $g^{(2)}_{\alpha\alpha}(\tau =0)\approx 2$ is enhanced nonetheless, due to a stimulated emission effect, which ultimately also originates in the existence of two-photon excitations.)


Considering now times, $\kappa_\alpha \tau \sim 1$, there will be contributions to the autocorrelation from an original state (before the first detection) with  double occupation of cavity $\alpha$  which decay. However, contributions can also stem from observation of a first decay from a single-occupied cavity $\alpha$,  followed by refilling of the cavity  and a consecutive second decay. These grow with time, as the refilling process takes some time. On short times, $\kappa_\alpha \tau \lesssim 1$, theory finds that the different time dependences of the two contributions tend to cancel each other, so that the autocorrelation remains roughly constant before finally exponential decay sets in.

This difference in time-dependence causes the detector-induced averaging to rather strongly suppress the cross-correlation, while it has a smaller effect on autocorrelations and thus moves the NRF away from 1/2.

\section*{Microwave chain}

\begin{figure*}[htb]
\centering
\includegraphics[width= \textwidth]{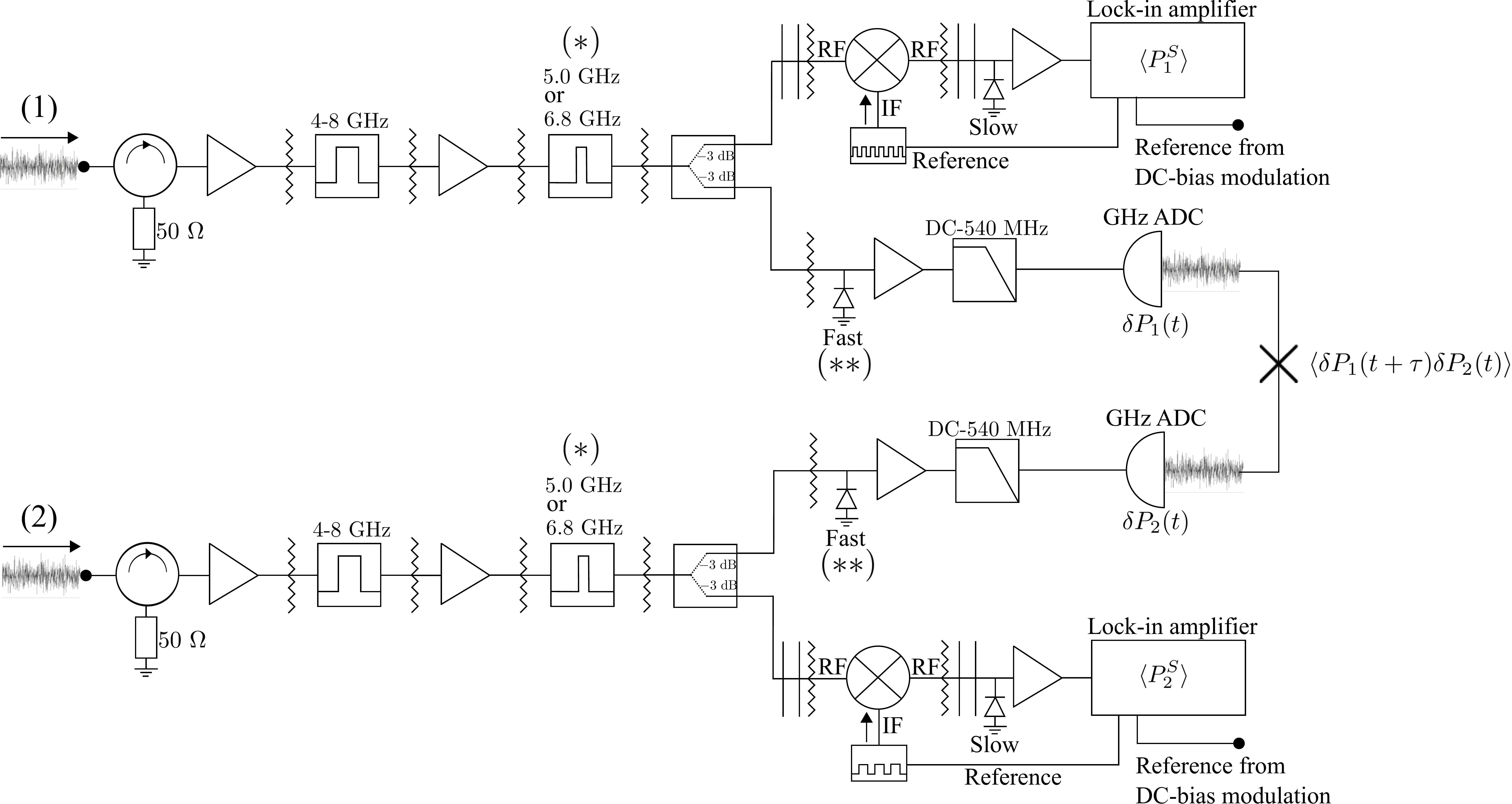}
\caption{\label{fig:correlationsetup}Circuit scheme of our room-temperature power correlation measurement chain detailing the 
300~K part of our experiment shown in Fig.~\ref{fig:scheme}(b) of the main text. The chain consists of two equal measurement 
lines '1' and '2' permitting us to perform cross-correlation measurements of the power emitted by the sample, entering the 
two individual lines.}
\end{figure*}

The purpose of this section is to describe in detail, first, the basic functionality of our correlation 
setup which we use to directly measure the quantities 
$\left\langle P_1^S\right\rangle$, $\left\langle P_2^S\right\rangle$ and the correlator 
$\left\langle \delta P_1(t+\tau)\delta P_2(t)\right\rangle$ to construct Eq.~(\ref{eq:g2exp}) of the main text 
without for every frequency combination ('aa', 'bb' and 'ab'). The basic idea is to \emph{assume} that the noise of the cryogenic amplifiers can be described as a thermal noise characterized by a constant noise temperature, uniform over modes $a$ and $b$, used as a reference to calculate the gains of the various elements of the detection chain. 

The power emitted by the sample is split in 
equal parts between the 3 dB hybrid coupler outputs shown in Fig.~\ref{fig:scheme}(b) of the main text. 
Hence, the output of each of the two branches of the beam splitter contains half of the 
power peaking at the two different frequencies $\simeq 4.9$ and $\simeq 6.7$~GHz of our resonators 
attached to the Josephson junction. The signals at the two outputs of the hybrid coupler is then fed into two cryogenic amplifiers sitting at 4.2K, as discussed in the main text and shown in  figure \ref{fig:scheme}-b. 
As shown by Fig.~\ref{fig:correlationsetup}, the output signals of each of the cryogenic amplifiers goes through  
isolator after which it is post-amplified with a conventional low-noise room-temperature amplifier, 
filtered through a 4-8~GHz bandpass filter and then, subsequently, it is once more amplified.
We further narrow the bandwidth of the detected signal with passive bandpass filters (labeled $(*)$) centered around $\nu_a$ and $\nu_b$, for subsequent correlation 
measurements: In order to measure the second order 
coherence function $g^{(2)}_{aa}(\tau)$ we use passive bandpass filters in each of 
the two measurement lines centered around 5~GHz with a bandwidth of 700~MHz 
in line '1' and 655~MHz in line '2' and for the measurement of $g^{(2)}_{bb}(\tau)$ 
we use bandpass filters centered around 6.8~GHz with a bandwidth of 650~MHz in 
line '1' and 660~MHz in line '2'. For the measurement of $g^{(2)}_{ab}(\tau)$ we use 
a bandpass filter centered around 6.8~GHz with a bandwidth of 650~MHz build into 
line '1' and a bandpass filter centered around 5~GHz with a bandwidth of 655~MHz 
build into line '2' (note that our results for the second order coherence functions 
are invariant against changing the filtering from one line to the other which we 
verified in an extra measurement run).

A power divider (-3dB splitter) distributes half of the bandpass filtered power into a 
measurement circuit which measures the mean power associated with the outgoing photon 
mode $\langle P_{1}^{S}\rangle$ or $\langle P_{2}^{S}\rangle$, measured by the respective 
measurement line '1' and '2'. The other half of the power is send into a different circuit which measures
the power fluctuations around the mean value of the total power coming from the cryogenic amplifiers, 
i.e.~$\delta P_{1,2}(t) = P_{1,2}(t)-\langle P_{1,2}\rangle$. 

\section*{Mean power measurements}

The measurement of the mean output power of each microwave chain is made using a 'slow' quadratic detector (with a response time in the $\mu$s range), whose output voltage is proportional to the microwave power at its input. We use a double lock-in technique to measure both the average power emitter by the sample $\langle P_{1,2}^{S}\rangle$ and the
total power $\langle P_{1,2}\rangle$ of our measurement chain, mainly dominated by the noise of the 
cryogenic amplifiers at 4.2~K. This last point will allow us to compensate for slow fluctuations of the gain of the chain, still assuming that the noise temperature of the cryogenic amplifiers is constant. 

\textbf{Measurement of the total average power $\langle P_{1,2}\rangle$}: The modulation is performed with a mixer circuit which we use as a 
switch, the switching is performed via the intermediate frequency (IF) port of the mixer using an attached
square wave generator operating between 0V (no power transmitted through the mixer) and 1.1 V (power transmitted through the mixer) at 113 Hz, and we detect the resulting square wave modulation of the output voltage of the quadratic detector by a standard Lock-in technique, after a final low-frequency amplification. The in- and output of the mixer ('switch') circuit is connected through DC-blocks 
(straight vertical lines in Fig.~\ref{fig:correlationsetup}) to the measurement 
chain to prevent possible near-DC noise from the modulation sources to disturb our low-noise measurements.
Furthermore, at each connection 
between microwave components we add attenuators (wiggly vertical lines) 
to flatten out standing waves due to possible imperfect SMA connections and 
to ensure a linear response of the complete measurement chain.

\textbf{Measurement of the power emitted by the sample $\langle P_{1,2}\rangle$}: Second, the excess power emitted by the sample when a voltage bias is applied 
appears on top of the noise floor of our measurement chain, described previously, 
and is measured in the following way. By performing simultaneous 
to the signal modulation for the total power measurement, a modulation of the sample bias 
voltage at a different reference frequency, we can separately measure the excess power $\langle dP\rangle$ 
coming exclusively from the sample by performing an additional lock-in detection of the power 
at the frequency of the bias voltage modulation. This measurement is performed on the same 
measurement branch which we use to measure the total power $\langle P \rangle$ of our 
measurement chains.

The mean power associated with the outgoing photon mode is then given by the following equation
\begin{equation}
\label{eq:meanP}
\langle P_{1,2}^{S} \rangle = \frac{\langle dP_{1,2} \rangle}{\langle P_{1,2} \rangle} k_{B} T_{N1,N2} \Delta \nu_{1,2}~,
\end{equation}
where $\langle dP_{1,2} \rangle$ indicates the mean of the excess power emitted by the sample 
into the measurement chains '1' and '2', respectively. Furthermore, $T_{N1,N2}$ are the noise 
temperatures of the cryogenic microwave amplifiers refered to the output of the sample 
(their values depend on whether line '1' or '2' is read out 
and on the detected mode 'a' or 'b') and $\Delta \nu_{1,2}$ is the bandwidth of our passive 
filtering to select the power emerging from modes 'a' and 'b'.

 \section*{Characterization of the two microwave resonators}

The resonators are patterned in a Niobium layer of d = 150~nm 
sputtered on a Silicon wafer covered by a 520~nm oxide layer. Both resonators consist of 
three quarter wavelength sections, one with an impedance slightly higher than $50~\Omega$, to increase the impedance 
above $50~\Omega$, the second with a low impedance to reach an impedance of a few Ohms, and a last section with an 
impedance above $100~\Omega$, to reach an impedance in the kilo-Ohms range. Table~\ref{tab:app01} contains the measured 
dimensions, where adopt the usual notations: $W$ stands for the width of the inner conductor, $W$ for the gap between the inner conductor and the ground plane of the CPW and $L$ is the length of the quarter wavelength section. Note that all sections are not designed exactly of identical length, in order to compensate for finite thickness effects of the Nb layer and of the oxide layer, and get the fundamental frequencies of the three sections stacked on one side of the junction identical. However, as shown later, we mistakenly neglected the kinetic inductance of the Nb, which made the resonant frequency of the last quarter wavelength section (i.e. the closest to the junction) significantly lower than the two other ones.

\vskip 0.5cm
\begin{table*}[htb]
\centering
\caption{\label{tab:app01}Measured CPW dimensions to calculate $Z(\nu)$.}
\begin{tabular}{|c|c|c|}
 \hline
   & Resonator A & Resonator B \\
  \hline
  First Section & $S=23.1 \mu$m  $W=38.5 \mu$m $L=$5.81 mm & $S=13 \mu$m  $W=43.5 \mu$m $L=$4.23 mm \\
  Second Section  & $S=95 \mu$m  $W=2.5 \mu$m $L=$5.81 mm & $S=95 \mu$m  $W=2.5 \mu$m $L=$4.23 mm \\
	Third Section  & $S=1.33 \mu$m  $W=49.3 \mu$m $L=$5.81 mm & $S=1.33 \mu$m  $W=49.3 \mu$m  $L=$4.23 mm\\
  \hline
\end{tabular}
\end{table*}
\vskip 0.5cm
The low temperature resistance of the Niobium was measured to be 3.8 k$\Omega$ for a critical temperature of 8~K. 
The low temperature resistivity is thus $\rho=$69 n$\Omega$m$^{-1}$. The London penetration length is thus 
\begin{equation}
\lambda=\sqrt{\frac	{\hbar \rho}{\mu_0 \pi\Delta}}= 120~\mathrm{nm},
\end{equation}
yielding a surface inductance
\begin{equation}
L_s=\mu_0\lambda \mathrm{cotanh} \frac{d}{\lambda}=1.8 \ 10^{-13}~\mathrm{H}.
\end{equation}
The additional kinetic inductance reads 
\begin{equation}
L_K=\frac{L_s}{4 S (1-k^2)K^2(k)}\left[\pi + \ln\left(\frac{4 \pi S}{d}\right) - k \ln\left(\frac{1+k}{1-k}\right)\right],
\end{equation}
where $k=\displaystyle{\frac{S}{S+2W}}$, and $K$ is the complete elliptic integral of the first kind.  
One thus obtains a kinetic inductance of $L_K=1.44 \ 10^{-7} \mathrm{H \ m}^{-1}$ for the narrowest sections 
and negligible for the others. This is to be compared with the $L_0=11.0 \ 10^{-7} \mathrm{H \ m}^{-1}$ electromagnetic 
inductance. This should decrease the resonant frequency by 6\%. In addition, the finite thickness of Niobium (not taken 
into account when designing the experiment) increases the wave velocity by 2\%, and the lower dielectric constant of silicon 
oxide increases the wave velocity and the impedance by 0.7\%. Hence in total we summarize our results in Table~\ref{tab:app02}.
\vskip 0.5cm
\begin{table*}[htb]
\centering
\caption{\label{tab:app02}Impedance and fundamental mode of the three quarter wavelength sections of each resonator.}
\begin{tabular}{|c|c|c|}
 \hline
   & Resonator A & Resonator B \\
  \hline
  First Section & $Z=67.5 \Omega$ $f_0= 5.14$ GHz & $Z=67.5 \Omega$ $f_0= 7.05$ GHz  \\
  Second Section  & $Z=23.1 \Omega$ $f_0= 5.18$ GHz & $Z=23.1 \Omega$ $f_0= 7.11$ GHz \\
	Third Section & $Z=142 \Omega$ $f_0= 4.9$ GHz & $Z=142 \Omega$ $f_0= 6.73$ GHz \\
  \hline
\end{tabular}
\end{table*}
\vskip 0.5cm

\section*{Comparison with high bias shot noise data}

We perform the noise temperature calibration by using our Josephson junction as a 
shot noise source connected to a known frequency dependent electromagnetic 
environment of impedance $Z(\nu)$, made of coplanar waveguide (CPW) resonators with 
known dimensions. The difference in power which is deposited in the electromagnetic environment
when biasing the Josephson junction at two different voltages 
$V_{1} = 0.975$~mV and $V_{2} = 0.762$~mV has then to fulfill the following equation:
\begin{widetext}
\begin{equation}
\label{eq:calibTN}
k_{B} T_{N} \frac{1}{\mathcal{B}}\int_{\nu_{min}}^{\nu_{max}}d\nu 
\left(\frac{\langle dP \rangle}{\langle P \rangle}(\nu)\bigg\rvert_{V_{1}} 
- \frac{\langle dP \rangle}{\langle P \rangle}(\nu)\bigg\rvert_{V_{2}}\right)
=
\frac{eV_{1} - eV_{2}}{2} \frac{1}{\mathcal{B}} 
\int_{\nu_{min}}^{\nu_{max}}d\nu~4\frac{\mathrm{Re}\left[Z(\nu)\right] R_{N}}{\lvert Z(\nu) + R_{N} \rvert^{2}}~.
\end{equation}
\end{widetext}
Here, the right-hand side of the equation quantifies the difference in noise power coupling of the Josephson 
junction with normal-state tunnel resistance $R_{N}$ to the electromagnetic 
environment having a complex valued impedance $Z(\nu)$ when the two voltages $V_{1}$ and $V_{2}$ 
are applied. The left-hand side resembles 
again Eq.~(\ref{eq:meanP}), but this time also for the difference in the power 
associated to the outgoing photon modes when the 
two voltages $V_{1}$ and $V_{2}$ are applied.
\begin{figure}[tb]
\centering
\includegraphics[width= \columnwidth]{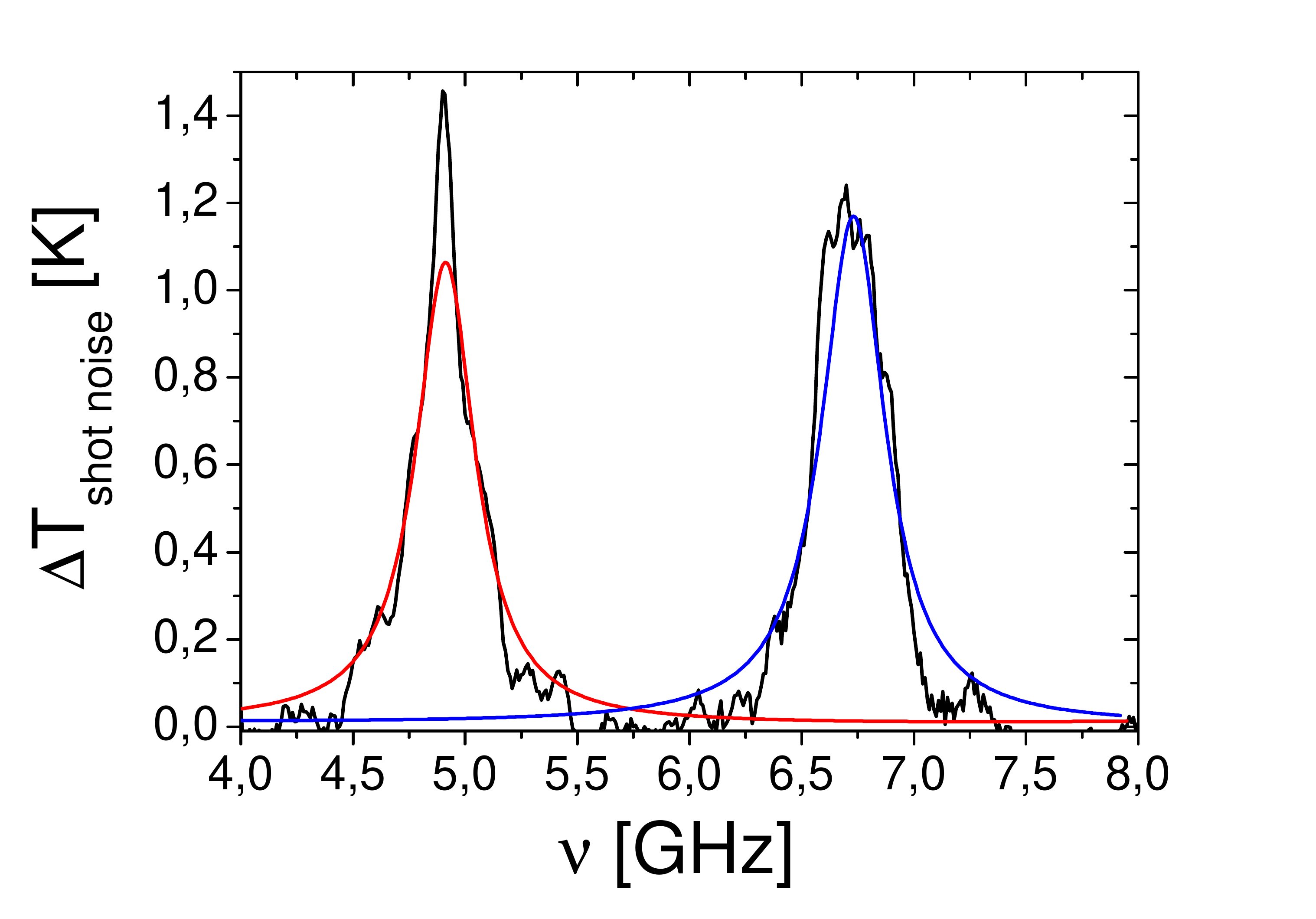}
\caption{\label{fig:excessT}Verification of the signal noise temperature calibration in 
the shot noise regime. The black line shows the experimentally determined 
difference in signal noise temperature (excess noise) when the sample is biased at 
$V = 0.975$~mV and $V = 0.762$~mV. The red and blue line is the theoretical 
comparison with the same bias parameters as input value and including the modeling of our 
electromagnetic environment. Note that only for the 
presentation of this figure we averaged $T_{N}$ for equal modes of the two distinct 
measurement lines '1' and '2', justified due to their match with an uncertainty of only 5\% or better.}
\end{figure}
The full equation is evaluated in the following way to calibrate $T_{N}$ with respect to 
the output of the two CPW resonators for each measurement line '1' and '2' and for 
each mode 'a' and 'b'. We evaluate the integrals within a bandwidth 
$\mathcal{B} = \nu_{max} - \nu_{min}$ which covers the two peaks in 
$\mathrm{Re}\left[Z(\nu)\right]$ around $\simeq 4.9$ and $\simeq 6.7$~GHz, 
whereas $\mathcal{B} \approx 700$~MHz which is to good approximation the 
bandwidth of all four passive bandpass filters we have employed for the correlation 
measurements. In this bandwidth we assume that $T_{N}$ does not change 
significantly so that we can pull this term out of the left integral in Eq.~(\ref{eq:calibTN}). 
Then we finally obtain four noise temperatures quantifying $T_{N}$ 
for each measurement line and mode. We then verify our calibration by taking our 
shot noise data and calculate the signal noise temperatures
$T_{N} \langle dP \rangle / \langle P \rangle$ for the same mode but for 
the two distinct measurement lines '1' and '2'. We find that the two calibrations match 
with an uncertainty of only 5\% or better. 
Figure~\ref{fig:excessT} finally summarizes our calibration procedure and 
shows the non-integrated equation (\ref{eq:calibTN}), suggesting that our assumption of a constant 
noise temperature within a 700~MHz bandwidth around $\simeq 4.9$ and $\simeq 6.7$~GHz is 
well justified.

\section*{Power fluctuations measurement}

We first measure the power emitted by the sample $\langle P_{1,2}^{S} \rangle$ using a square wave modulation on the voltage across the Josephson junction, as discussed above. 
After a few seconds the measurement of $\langle P_{1,2}^{S} \rangle$ is finished and we measure 
the correlator $\left\langle \delta P_1(t+\tau)\delta P_2(t)\right\rangle$ which takes another few seconds.
We verify that within this short measurement time the output power of the experiment is stable. 
For the measurement of the power-power fluctuations correlator we switch off the bias voltage modulation, previously used to measure 
$\langle dP\rangle$ with the lock-in amplifier, and bias the Josephson junction now with a constant voltage.

Power fluctuations are measured by 
the measurement circuit behind the -3dB splitter represented in Fig.\ref{fig:correlationsetup} not used for the average power measurement. This circuit consists of a fast 
quadratic detector of type 'Herotek DTM 180AA' with a response time of 0.425~ns which we use to 
sample the power fluctuations. Its output voltage is amplified 
and low-pass filtered by a DC-540~MHz filter which connects to the input of a SP-Devices ADQ 412 fast 
acquisition card. In our measurements we use only two out of four ADCs on the acquisition card which 
provide us a maximum bundled sample rate of 2~Gigasamples/s or equivalently a time resolution of 
0.5~ns. In this paper we focus on the zero time second order coherence functions, i.e.~for which 
$\tau = 0$. First of all our microwave measurement setup is built in such a way to keep the physical 
microwave line length almost equal, having in mind that a difference in length of 
about 10~cm corresponds to a time shift of 0.5~ns. Further balancing of the physical line length is 
performed electronically by shifting the acquired time bins on the acquisition card. For 
this calibration we usually conduct a measurement of the time resolved shot noise correlation
at 4.9 or 6.7~GHz where we determine the exact position of the maximum of 
$\left\langle \delta P_1(t+\tau)\delta P_2(t)\right\rangle$, corresponding to $\tau = 0$ 
between the two measurement lines.

Since the acquisition card measures the incoming power fluctuations in units of \emph{square digits}, we need to 
translate them to a power quantity. To do so, we rely on the spectral density of the auto-correlated power fluctuations 
of the thermal noise of the cryogenic amplifiers, which reads

\begin{equation}
\begin{aligned}
S_{PiPi}(f)& = k_B^2 T_{ni}^2 \Delta \nu_i \left(1-\frac{f}{\Delta \nu_i}\right) & \mathrm{for}& \ \ 0 < f \le \Delta \nu_i \\
& = 0 &\mathrm{for}& \ \ f > \Delta \nu_i ~.
\end{aligned}
\end{equation}
Integrating over $0<f<\Delta \nu_i$, we get the expected auto-correlated output of chain $i$ of the ADC:

\begin{equation}
\mathcal{A_i} = G_i^2 k_B^2 T_{ni}^2 \Delta \nu_i^2,
\end{equation}
 
where $G_i$ is the total gain of chain $i$, in Digit/Watt. 

\begin{figure}[tb]
\centering
\includegraphics[width= \columnwidth]{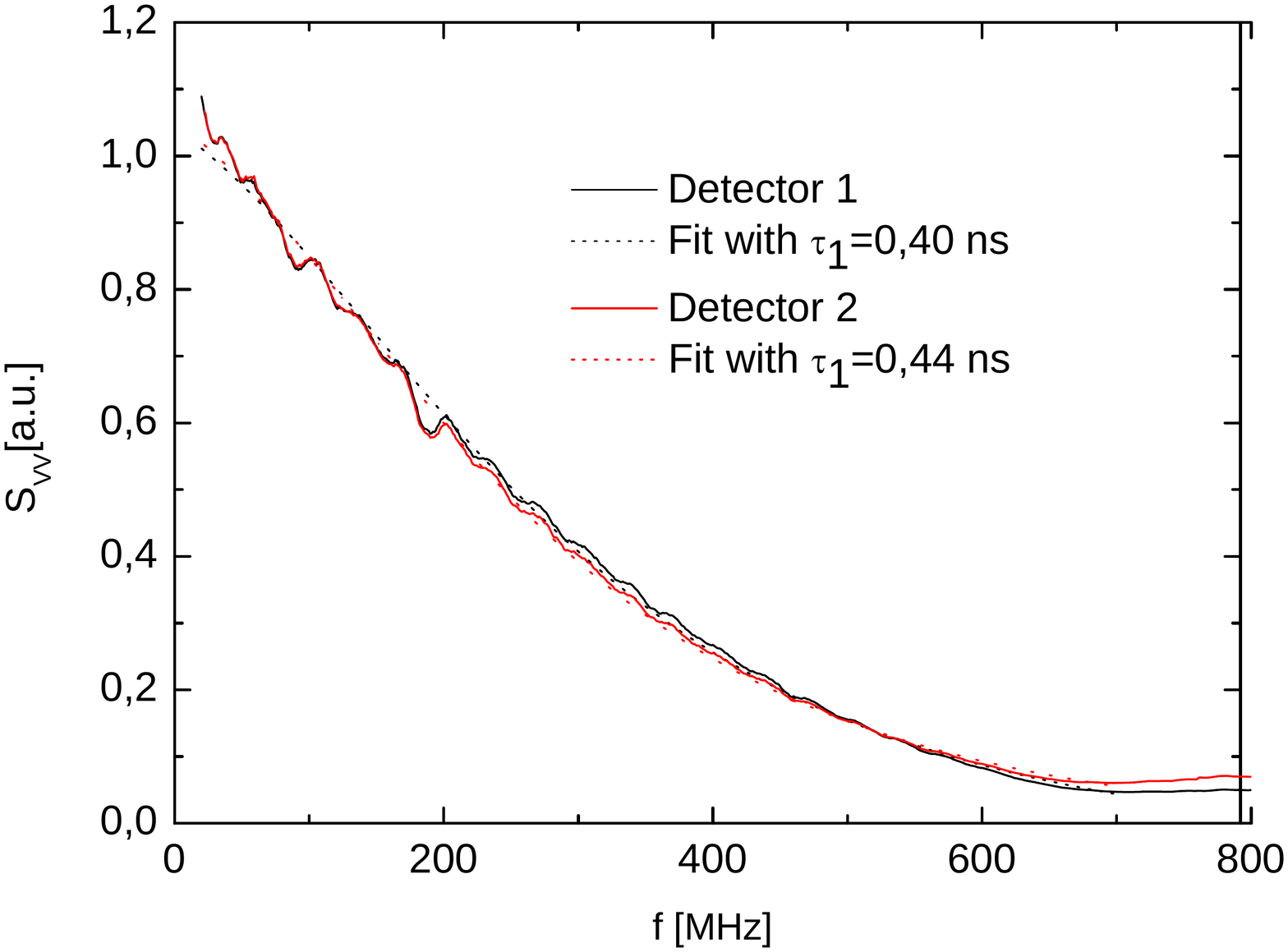}
\caption{\label{fig:detresp}Solid lines show the measured normalized noise power spectral density $S_{VV}(f)$ of the output voltage of our two 'Herotek DTM 180AA' fast quadratic detectors (parts labeled $(**)$ in Fig.~\ref{fig:correlationsetup}). Dotted lines are fit to the behavior of an ideal detector, followed by a first order filter, yielding response times $\tau_1$=0.40 ns and $\tau_2$=0.44 ns.}
\end{figure}

However, the finite response time of our fast quadratic detectors introduces a first-order filtering of the detected power fluctuations  with a certain cut-off frequency $(2\pi \tau_{det})$, where $\tau_{det}$ is the 
characteristic response time of the detector. 
We determine this response time we determine experimentally by injecting 
noise from our amplifier chain in a 700 MHz bandwidth into the input of the quadratic detector and by 
measuring its noise power spectral density, shown in Fig.~\ref{fig:detresp}. Fitting the experimental spectral density of the output voltage with 

 \begin{equation*}
\begin{aligned}
S_{ViVi}(f)& = A \frac{1-f/\Delta \nu_i}{1+ (2\pi f \tau_{det})^2} \ & \mathrm{for}& \ \ 0 < f \le \Delta \nu_i \\
& = 0 &\mathrm{for}& \ \ f > \Delta \nu_i ~
\end{aligned}
\end{equation*}
\noindent 
yields a response time of $\tau_{det} =0.40$~ns for detector 1 and $\tau_{det} =0.44$~ns for detector 2. This will reduce the  
detected auto-correlated fluctuations of the ADQ

\begin{equation}
\mathcal{A}_i = C_i G_i^2 k_B^2 T_{ni}^2 \Delta \nu_i^2,
\end{equation}
 
with $C_i$ ranging from 0.72 to 0.74 depending on the chosen filters and quadractic detectors. This allows us to determine $G-i$ and to express the correlator 
$\left\langle \delta P_1(t+\tau)\delta P_2(t)\right\rangle$ as a function of the cross-correlated and auto-correlated  fluctuations of the output of the ADQ:
\begin{equation}
\label{eq:correlatorP}
\left\langle \delta P_1(t+\tau)\delta P_2(t)\right\rangle = \sqrt{C_1 C2} \frac{\mathcal{C}}{\sqrt{\mathcal{A}_1 \cdot \mathcal{A}_2}}
k_{B}^2 T_{N1} T_{N2} \Delta \nu_1 \Delta \nu_2~.
\end{equation}
Here, $\mathcal{C}$, $\mathcal{A}_1$ and $\mathcal{A}_2$ denote the cross-correlation 
between lines '1' and '2' when a bias voltage is applied to the Josephson junction 
and the auto-correlations of line '1' and '2' when zero bias is applied to 
the Josephson junction, all three quantities are in units of \emph{square digits} as 
measured directly by the acquisition card.

In total, by substituting Eqs.~(\ref{eq:meanP}) and (\ref{eq:correlatorP}) into 
Eq.~(\ref{eq:g2exp}) of the main text the second order coherence function can 
be entirely expressed in terms of experimental quantities as:
\begin{equation}
\label{eq:g2exp02}
g^{(2)}_{\alpha,\beta}(\tau) = 1 + \sqrt{C_1 C_2} \frac{\mathcal{C}}{\sqrt{\mathcal{A}_1 \cdot \mathcal{A}_2}} 
\frac{\langle P_{1}\rangle \langle P_{2}\rangle }{\langle dP_{1}\rangle \langle dP_{2}\rangle}~.
\end{equation}

\begin{thebibliography}{31}
\bibitem[{Not({\natexlab{a}})}]{NoteEJ}
\bibinfo{note}{More specifically, to keep stimulated emission effects
  negligible, one must ensure that the number of photons is well below
  $1/\Delta_{a,b}^2$ with $\Delta_{a,b}=(2e^2 Z_{a,b}/\hbar)^{1/2}$.}
\bibitem[{Not({\natexlab{b}})}]{NoteQDtimeres}
\bibinfo{note}{We determine the response time of the quadractic detectors by
  measuring the spectral density of the fluctuations of their output voltage as
  a consequence of the fluctuations of the noise power of the amplifiers. More
  details on this point can be found in the SM.}
\end{thebibliography}
\end{document}